# Maximum capacity path problem with loss factors

Javad Tayyebi[1][a], Mihai-Lucian Rîtan[2][b] and Adrian Marius Deaconu[2][c]

[1]Department of Industrial Engineering, Birjand University of Technology, Industry and Mining Boulevard, Ibn Hesam Square, Birjand, Iran
[2]Department of Mathematics and Computer Science, Transilvania University of Brașov, Iuliu Maniu st. 50, Brașov, Romania
javadtayyebi@birjandut.ac.ir, {mihai.ritan, a.deaconu}@unitbv.ro

Keywords: Capacity path problems, combinatorial optimization, polynomial algorithms.

Abstract: The maximum capacity path problem is to find a path from a source to a sink which has the maximum capacity among all paths. This paper addresses an extension of this problem which considers loss factors. It is called the generalized maximum capacity path problem. The problem is a network flow optimization problem whose network contains capacities as well as loss factors for arcs. The aim of the problem is to find a path from an origin to a destination so as to send a maximum flow along the path considering loss factors and respecting capacity constraints. The paper presents a zero-one formulation of the problem and moreover, it presents two efficient algorithms which solve the problem in polynomial time.

## 1 INTRODUCTION

Combinatorial optimization is a special class of mathematical program that consists of finding an optimal object among a finite set of specific-structured objects. Some most prominent problems of this class are shortest path (SP) problems, maximum reliability path (MRP) problems, and maximum capacity path (MCP) problems. In these problems, the goal is to find an optimal path from an origin to a destination under a special objective function as follows:

1. $\min_{P \in \mathbb{P}} \sum_{(i,j) \in P} l_{ij}$ for SP problems
2. $\max_{P \in \mathbb{P}} \prod_{(i,j) \in P} p_{ij}$ for MRP problems
3. $\max_{P \in \mathbb{P}} \min_{P \in \mathbb{P}} u_{ij}$ for MCP problems

where $\mathbb{P}$ is the set consisting of all paths from the origin to the destination, $l_{ij}$, $p_{ij}$, and $u_{ij}$ denote respectively the length, the reliability, and the capacity of arc $(i,j)$. Fortunately, these problems are tractable, i.e., there are polynomial-time algorithms to solve them. Specially, shortest path problems can be solved by a Fibonacci-heap implementation of Dijkstra's algorithm in $O(m + n \log(n))$ if lengths are nonnegative; otherwise, the best-known algorithm is a FIFO implementation of Bellman-Ford algorithm which has a complexity of $O(mn)$, where n and m are the number of nodes and arcs, respectively. The maximum reliability path problem can be converted into a shortest path problem by setting $l_{ij} = -\log(p_{ij})$ for every arc $(i,j)$. So, it can be solved as well as the SP problem, especially in the case that $p_{ij} < 1$. Moreover, the both MRP and MCP problems can be solved directly by modifying the shortest path algorithms because they enjoy the optimality conditions similar with that of the SP problem (See (Ahuja, 1988) for more details). However, the best-known algorithm for solving the maximum capacity path problem in undirected network is not based on this concept, but it is a recursive algorithm with a linear complexity $O(m)$ (Punnen, 1991).

As an application of maximum capacity path problems, consider a network that represents connections between routers in the Internet. The capacity of an arc represents the bandwidth of the corresponding connection between two routers, the maximum capacity path problem is to find the path between two Internet nodes that has the maximum possible bandwidth. Besides this well-known network routing problem, MCP is also an important component of the Schulze method for deciding the winner of a multiway election (Schulze, 2011). It is applied to digital compositing (Fernandez, 1998), and metabolic pathway analysis (Ullah, 2009).

---

[a] https://orcid.org/0000-0002-7559-3870
[b] https://orcid.org/0009-0007-4601-6533
[c] https://orcid.org/0000-0002-1070-1383

This paper introduces a novel combinatorial optimization problem, called the generalized maximum capacity path (GMCP) problem. It is a more complex version of the problem of finding the directed path from a given source node s to a given sink node t that has the minimum loss among all directed paths from s to t (Deaconu, 2023). GMCP is defined on a network, which each arc has two attributes: capacity and loss factors. The capacity of an arc means the maximum amount that can flow on the arc. The loss factor of an arc is the flow value arriving at its tail node if 1 unit of flow sends on the arc. The generalized maximum capacity problem is to find a path which is capable of sending maximum flow considering loss factors. This problem is inspired from an extension of maximum flow problems regarding loss factors, called the generalized maximum flow problem [5]. So, its algorithms can be used as subroutines for solving generalized maximum flow problems. On the other hand, the GMCP problem is an extension of MRP and MCP problems because it is converted to a MRP problem in the case that capacities are infinite, and is transformed into a MCP problem if loss factors are equal to 1.

## 2 PRELIMINARIES

Let $G = (V, A, \boldsymbol{u})$ be a directed and connected network, where $V$ is the set of nodes, $A$ is the set of arcs which is a subset of $A \times A$ (an arc $a = (i,j)$ starts from node $i$ and terminates at node $j$), $\boldsymbol{u}: A \to \mathbb{R}^+$ is the capacity function, $s$ and $t$ are two special nodes of $V$, $s$ is called the source node and $t$ is referred to as the sink node. We denote by $n$ the number of nodes and by $m$ the number of arcs, i.e.,

$$n = |N| \text{ and } m = |A|.$$

A path $P$ from a node $w \in A$ to a node $v \in A$ in the network G is given by a sequence of nodes: $P: (w = i_1, i_2, \dots, i_l = v)$, where $l \geq 1$, and $(i_k, i_{k+1}) \in A$, for every $k = 1, 2, \dots, l-1$. For simplifying, hereafter, a path from $s$ to $t$ is called an $st$- path.

The capacity of an $st$- path $P$ is denoted by $u(P)$ and is given by the minimum of its capacities, that is,

$$u(P) = \min \{u(a) | a \in P\}.$$

The maximum capacity path problem (MCP) in the network G is to find an $st$-path $\tilde{P}$ having the maximum capacity among all $st$-paths:

$$u(\tilde{P}) = \max\{u(P) | P \text{ is an } s - t \text{ path}\}.$$

This problem is also called the widest path problem, the bottleneck shortest path problem, and the max-min path problem.

## 3 PROBLEM FORMULATION

This section focuses on the GMCP problem. Let us formally define the problem. Suppose that a connected and directed network $G(V, A, \boldsymbol{u}, \boldsymbol{p})$. Each arc $(i,j) \in A$ is associated to two parameters $u_{ij} \geq 0$ and $p_{ij} \in (0,1]$ which are referred to as the capacity and the loss factor, respectively. $u_{ij}$ is the maximum amount of flow which can be sent along $(i,j)$. The loss factors represent physical transformations due to evaporation, energy dissipation, breeding, theft, or interest rates (Tayyebi, 2019). If $x_{ij}$ units of flow enter arc $(i,j)$, then $p_{ij}x_{ij}$ arrive at $j$. So, arc $(i,j)$ absorbs $(1 - p_{ij})x_{ij}$ units of flow. The generalized maximum capacity problem is to find a path from the source node $s$ to the sink node $t$, referred to as an $st$-path, so that we can send the maximum possible flow along it considering loss factors. Let $x_{ij}$ be flow entering arc $(i,j)$ and $y_{ij}$ be a zero-one variable determining whether $(i,j)$ has a positive flow or not. The problem can be formulated as a mixed zero-one linear programming as follows:

$$\max z = v_t \tag{1a}$$

$$\sum_{j:(i,j)\in A} x_{ij} - \sum_{j:(j,i)\in A} p_{ji} x_{ji} \tag{1b}$$

$$= \begin{cases} v_s & i = s, \\ 0 & i \neq s, t, \forall i \in V, \\ -v_t & i = t, \end{cases}$$

$$\sum_{j:(i,j)\in A} y_{ij} \leq 1, \quad \forall i \in V \setminus \{t\}, \tag{1c}$$

$$0 \leq x_{ij} \leq u_{ij} y_{ij}, \quad \forall (i,j) \in A, \tag{1d}$$

in which $v_s$ is the flow leaving $s$, while $v_t$ is the flow entering $t$. Constraints (1b) and (1d) are respectively the same balanced and bound constraints of maximum flow problems (Ahuja,1988). Constraint (1c) states at most one leaving arc of any node is capable of sending flow. So, it guarantees that flow is sent only along one $st$-path.

Formulation (1) of GMCP problems is similar to that of generalized maximum flow problems, except that it contains zero-one variables $y_{ij}$ as well as the constraints (1c) (Ahuja, 1988).

**Remark 1.** Although we have assumed that $p_{ij} \leq 1$, constraint (1d) does not regard the probable flow increment on $(i, j)$ in the case that $p_{ij} > 1$. In this case, it must be written as $0 \leq \max\{x_{ij}, p_{ij}x_{ij}\} \leq u_{ij}y_{ij}$. Without any loss of generality, we can redefine the capacity of $(i, j)$ to be $\min\{u_{ij}, u_{ij}/p_{ij}\}$ for covering this situation.

## 3 ALGORITHMS

In this section, we develop two algorithms which solve the GMCP problem in polynomial time. This proves that this problem is tractable like to the MCP, SP, and MRP problems.

A simple observation is that if we send maximum flow along a path, then at least one of its arcs is saturated. The capacity of this arc determines the value of flow along the path. Suppose that $(i_P, j_P)$ be the last saturated arc belonging to a $st$-ath $P = s - \cdots - i_P - j_P - \cdots - k - t$. So, the flow value along $P$ is equal to $u_{i_P j_P} p_{i_P j_P} \ldots p_{kt}$. It is remarkable that if we remove the arc $(i_P, j_P)$ from the network and add an arc $(s, j_P)$ with capacity $u_{sj_P} = u_{i_P j_P}$ and loss factor $p_{sj_P} = p_{i_P j_P}$, then we can send the same flow value along the new path $s - j_P - \cdots - k - t$. This simple idea yields a polynomial-time algorithm to solve problem (1). In the first algorithm, without considering arc capacities, we find a maximum reliability path from $s$ to $t$, namely, a path $P$ for which the value $\prod_{(i,j) \in P} p_{ij}$ is maximized. This can be performed by defining arc lengths $l_{ij} = -log(p_{ij})$ and finding a shortest $st$-path with respect to arc lengths $l_{ij}$. Let $P$ be found. Then, we find the last arc $(i_P, j_P)$ of $P$ to be saturated if we send maximum flow along $P$. We remove the arc $(i_P, j_P)$ from the network and add an artificial arc $(s, j_P)$ with factor $p_{sj_P} = u_{i_P j_P} p_{i_P j_P}$. Then, we repeat the procedure until a path $P$ is found so that the last saturated arc is one of artificial arcs. Notice that the optimal value of the problem (1) is equal to the maximum flow along the last path found by the algorithm. To obtain the optimal path, it is required that we save the portion of path $P$ from $s$ to $i_P$ whenever we remove $(i_P, j_P)$ and add $(s, j_P)$. This can be done by introducing an extra parameter $P_{sj_P}$ for each artificial arc. So, if the algorithm finds $P$ in last iteration, then the optimal solution is a path passing arcs of $P_{sj_P} \cup P\setminus\{(s, j_P)\}$. You can see a formal description of our first algorithm in Algorithm 1. Since an arc is removed in each iteration, it follows that the number of iterations is at most equal to $m$. So, we have the following result.

**Theorem 1:** The complexity of Algorithm 1 is $O(mS(n, m))$ in which $S(m, n)$ is the complexity of finding a shortest path in the network.

---
**Algorithm 1**

---
Input: An instance of the generalized MCP problem
Output: An optimal path

**for** $(i, j) \in A$:
  **if** $i == s$:
    Set $\bar{l}_{ij} = -\log p_{ij} u_{ij}$ and $P_{ij} = (i, j)$
  **else**:
    Set $\bar{l}_{ij} = -\log p_{ij}$ and $P_{ij} = \emptyset$
**while** True:
  Find a shortest path $P$ with respect to $\bar{l}_{ij}$.
  **if** $P_{sj_P} \neq \emptyset$:
    The optimal path is $P_{sj_P} \cup P\setminus\{(s, j_P)\}$.
  **else:**
    Find the last arc $(i_P, j_P)$ of $P$ to be saturated.
    Remove $(i_P, j_P)$.
    Add an artificial arc $(s, j_P)$.
    Set $\bar{l}_{sj_P} = -\log(u_{i_P j_P} p_{i_P j_P})$.

---

Now, we present some optimality conditions for problem (1). This enables us to design an algorithm to solve the problem in $O(S(m, n))$ time. This will improve the complexity of Algorithm 1 by a factor $m$.

We first define a label $d(j)$ for every node $j \in V$. At intermediate stages of computation, this label $d(j)$ is an estimate of (an upper bound on) the maximum flow sent from the source node $s$ to node $j$ along one path, and at termination it is the optimal value of problem (1). We should provide necessary and sufficient conditions for a set of labels to represent maximum flow.

Let $d(j)$ for $j \neq s$ denote the value of maximum flow sent from the source node to the node $j$ [we set $d(s) = +\infty$]. If the labels are optimal, they must satisfy the following necessary optimality conditions:

$$d_j \geq p_{ij} \min\{u_{ij}, d_i\}.$$

This is an extension of the optimality conditions of both the MCP and MRP problems. On the optimal path, the inequality is satisfied in the equality form. It states that the label of node $j$ is either $p_{ij}d_i$ or $p_{ij}u_{ij}$. In the case that the flow value arrived at node $i$ is less than $u_{ij}$, (namely, $d_i < u_{ij}$), this arc is not saturated and consequently, $d_j = p_{ij}d_i$. In the other case, $(i,j)$ is saturated, and it interdict sending flow more than its capacity. So, $d_j = p_{ij}u_{ij}$ in this case.

Since this optimality condition is similar to that of the SP problem, we can apply the concept of Dijkstra's algorithm to solve problem 1. This is presented in Algorithm 2.

**Theorem 2:** Algorithm 2 solves the problem in $O(n^2)$ time.

*Proof.* Since each arc is check only one time in the for loop, it follows that the number of iterations of the two last lines is at most $O(m)$ ($\leq O(n^2)$). On the other hand, the node selection with minimum label requires $O(n)$ in each iteration. Since the number of iterations is $O(n^2)$, the bottleneck operation is the same node selection with $O(n^2)$ time.

---
**Algorithm 2**
---

Input: An instance of the generalized MCP problem
Output: An optimal path

**for** $i \in V$:
    Set $d(i) = 0$.
Set $d(s) = +\infty$.
Set $S = \emptyset; \bar{S} = V$;
**while** $|S| < n$:
    Let $i \in \bar{S}$ be a node for which
        $d(i) = \min\{d(j): j \in \bar{S}\}$;
    Update $S = S \cup \{i\}; \bar{S} = \bar{S}\setminus\{i\}$;
    **for** each $j \in V: (i,j) \in A$:
        **if** $d_j < p_{ij}\min\{u_{ij}, d_i\}$:
            Update $d_j = p_{ij}\min\{u_{ij}, d_i\}$

---

We can also use the available implementations of Dijkstra's algorithm for the complexity improvement of Algorithm 2. For example, the Fibonacci heap implementation reduces the complexity to $O(m + n\log n)$.

## 4 EXPERIMENTS & DISCUSSIONS

It is clear that Algorithm 2 is faster than Algorithm 1 (see Theorem 1 and Theorem 2). However, Algorithm 1 has the advantage that it can be efficiently parallelized on GPUs, since it successively runs classical Dijkstra's algorithm (Ortega-Arranz, 2013). Consequently, using CUDA implementation of Dijkstra's algorithm led to significant speed-up of Algorithm 1. As a result, CUDA version is faster (up to 25.7 times faster) than Algorithm 2 (see Table 2) on big instances of networks (10,000 nodes and more).

Table 1: Network generator parameters used for experiments.

| No. of nodes | No. of Instances | No. of paths | No. of cycles | Erdős–Rényi prob. | No. |
|---|---|---|---|---|---|
|  |  | 500 | 100 | 0.5 | 1 |
| 1000 | 10000 | 500 | 100 | 0.7 | 2 |
|  |  | 500 | 100 | 0.9 | 3 |
|  |  | 1000 | 100 | 0.1 | 4 |
| 2000 | 1000 | 1000 | 250 | 0.15 | 5 |
|  |  | 1000 | 500 | 0.6 | 6 |
|  |  | 2500 | 1000 | 0.1 | 7 |
| 5000 | 100 | 2500 | 1000 | 0.2 | 8 |
|  |  | 2500 | 1000 | 0.3 | 9 |
|  |  | 5000 | 2500 | 0.15 | 10 |
| 10000 | 5 | 5000 | 2500 | 0.3 | 11 |
|  |  | 5000 | 2500 | 0.5 | 12 |
| 15000 | 3 | 7500 | 1000 | 0.15 | 13 |
| 20000 | 2 | 7500 | 1000 | 0.15 | 14 |
| 25000 | 1 | 8000 | 1500 | 0.15 | 15 |

Table 2: Running times (ms) comparison between Algorithm 1 (CPU and GPU) and Algorithm 2 CPU.

| No. | Alg. 1 CPU | Alg. 1 GPU | Alg. 2 CPU | Alg. 1 GPU vs Alg. 2 CPU (times) |
|---|---|---|---|---|
| 1 | 165.75 | 187.30 | 18.32 | **0.10** |
| 2 | 121.1 | 138.66 | 8.24 | **0.06** |
| 3 | 188.96 | 311.78 | 14.84 | **0.05** |
| 4 | 30.22 | 52.40 | 6.86 | **0.13** |
| 5 | 120.00 | 209.34 | 63.00 | **0.30** |
| 6 | 193.50 | 315.60 | 19.50 | **0.06** |
| 7 | 183.00 | 71.52 | 40.15 | **0.56** |
| 8 | 284.30 | 103.41 | 41.7 | **0.40** |
| 9 | 626.6 | 226.74 | 61.4 | **0.27** |
| 10 | 671.10 | 89.21 | 98.01 | **1.10** |
| 11 | 677.14 | 67.78 | 216.00 | **3.19** |
| 12 | 940.20 | 76.15 | 358.01 | **4.70** |
| 13 | 1306.07 | 49.41 | 238.10 | **4.82** |
| 14 | 2965.04 | 53.48 | 452.11 | **8.46** |
| 15 | 3549.10 | 32.11 | 826.04 | **25.72** |

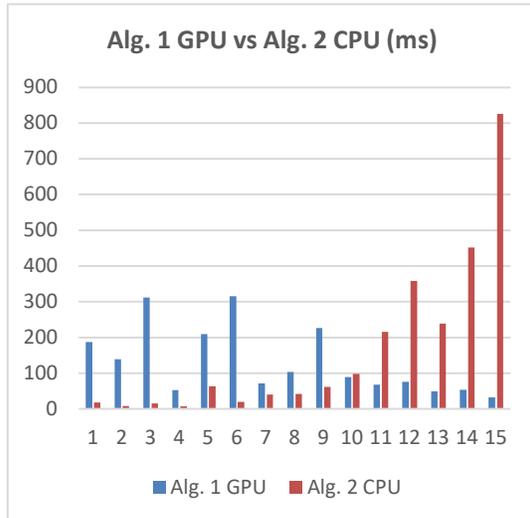

Figure 1: Running times (ms) comparison between Algorithm 1 GPU and Algorithm 2 CPU.

We used different instances of random generated networks having the number of nodes varying from 1000 to 25000 (see Table 1). The networks were generated using the network random generator from (Deaconu, 2021).

The experiments were performed using a PC with an Intel(R) Core(TM) i5-6500 CPU @ 3.20GHz, 24 GB RAM, and an NVIDIA GeForce GTX 1070 TI graphics card. The algorithms were programmed in Visual C++ 2022 under Windows 10.

## 5 CONCLUSIONS

In this paper a new combinatorial optimization problem was introduced. It is to find a path which is capable of sending maximum flow considering both, capacities and loss factors of the arcs. This problem is called the generalized maximum capacity path problem.

Two strongly polynomial algorithms were presented to deal with GMCP. The first one runs in $O(mS(n,m))$, where $S(m,n)$ is the time complexity of finding a shortest path in the network. The second one has a better time complexity of $O(n^2)$ time, but if the first one is implemented of GPUs, it runs significantly faster than the second one on big instances of networks.

## REFERENCES

Ahuja, R. K., Magnanti, T. L., & Orlin, J. B. (1993). *Network flows*, Prentice Hall.

Deaconu, A.M., & Spridon, D., (2021). *Adaptation of Random Binomial Graphs for Testing Network Flow Problems Algorithms*. Mathematics, 9(15), 1716.

Deaconu, A.M., Ciupala, L., & Spridon, D., (2023). *Finding minimum loss path in big networks*. 22nd International Symposium on Parallel and Distributed Computing (ISPDC), Bucharest, Romania, pp. 39-44.

Fernandez, E., Garfinkel, R., & Arbiol, R. (1998). *Mosaicking of aerial photographic maps via seams defined by bottleneck shortest paths*. Operations Research, 46(3), 293-304.

Ortega-Arranz, H., Torres, Y., Llanos, D.R., & Gonzalez-Escribano, A., (2013). *A new GPU-based approach to the Shortest Path problem*. 2013 International Conference on High Performance Computing & Simulation (HPCS), Helsinki, Finland, pp. 505-511.

Punnen, A. P. (1991). *A linear time algorithm for the maximum capacity path problem.* European Journal of Operational Research, 53(3), 402-404.

Schulze, M. (2011). *A new monotonic, clone-independent, reversal symmetric, and Condorcet-consistent single-winner election method.* Social choice and Welfare, 36, 267-303.

Tayyebi, J., & Deaconu, A. (2019). *Inverse generalized maximum flow problems.* Mathematics, 7(10), 899.

Ullah, E., Lee, K., & Hassoun, S. (2009, November). *An algorithm for identifying dominant-edge metabolic pathways*. In 2009 IEEE/ACM International Conference on Computer-Aided Design-Digest of Technical Papers (pp. 144-150). IEEE